\documentclass[submission,copyright,creativecommons]{eptcs}
\providecommand{\event}{Extended Version of Preprint submitted to FMAS 2021} 
\usepackage{breakurl}             
\usepackage{underscore}           
\usepackage{amssymb}
\usepackage{graphicx}
\usepackage{tabularx}
\usepackage[justification=centering]{subfig}
\usepackage{amsthm}
\newtheorem{lemma}{Lemma}
\theoremstyle{definition}
\newtheorem{definition}{Definition}
\title{Finding the Best Partitioning Policy for Efficient Verification of Autonomous Systems at Runtime}
\author{ Melika Dastranj \qquad Mehran Alidoost Nia \qquad Mehdi Kargahi
\institute{DRTS Lab, School of Electrical and Computer Engineering, College of Engineering, \\University of Tehran, Tehran, Iran}
\email{\quad \{melika.dastranj, alidoostnia, kargahi\}@ut.ac.ir}
}
\def\titlerunning{Finding the Best Partitioning Policy for Verification of Autonomous Systems}
\def\authorrunning{M. Dastranj, M. Alidoost Nia \& M. Kargahi}
\begin{document}
\maketitle
\begin{abstract}
The autonomous systems need to decide how to react to the changes at runtime efficiently. The ability to rigorously analyze the environment and the system together is theoretically possible by the model-driven approaches; however, the model size and timing limitations are two significant obstacles against such an autonomous decision-making process. To tackle this issue, the incremental approximation technique can be used to partition the model and only verify a partition if it is affected by the change. This paper proposes a policy-based analysis approach that finds the best partitioning policy among a set of available policies based on two proposed metrics, namely Balancing and Variation. The metrics quantitatively evaluate the generated components from the incremental approximation scheme according to their size and frequency. We investigate the validity of the approach both theoretically and experimentally via a case study on energy harvesting systems. The results confirm the effectiveness of the proposed approach. 
\end{abstract}

\section{Introduction}
Autonomous systems primarily work through uncertain environments, and due to the changes, they must autonomously decide about adaptive actions. To include all aspects of the system in decision-making, one can use a model-driven approach~\cite{model} to rigorously analyze the local system and the environment for changes and decide what action better adapts the autonomous system regarding formal specifications. The main obstacle against runtime analysis of autonomous systems is the size of the model that is relatively large for a resource-constraint system to meet timing limitations~\cite{limitation}. Therefore, we need to verify the system at runtime efficiently. 

The early solution to this problem is to use approximation~\cite{abate1, abate2}. As the model is changed at runtime, the approximation must be repeated, bringing overhead to the verification process. One can use incremental approximation~\cite{man1} to enhance the reusability of previously verified parts of the model~\cite{incremental}. In this approach, we partition the model into a set of independent components. At runtime, if a change occurs, we need to only re-approximate/re-verify those components that are affected by the changes. This improves the verification process under a certain upper bound for errors and a correct aggregation algorithm for a central decision-making configuration. 

According to the stochastic nature of the environment and the ability to choose an adaptive action in autonomous systems, we can deploy the Markov decision process (MDP) to model the system~\cite{man2}. The changes are brought to the model as a set of parameters. Then, we use a parametric MDP (pMDP)~\cite{pmdp} for capturing the changes. In this research, we use a case study on an energy-harvesting system that uses a MAPE-K loop for adaptation purposes~\cite{man3}. MAPE-K stands for monitoring, analyzing, planning, executing, and knowledge. In this research, we focus on the analyzing round of the MAPE-K loop. 

In this research, we tackle the problem of efficiently deciding about the changes in autonomous systems at runtime. We use a pMDP as the modeling construct and apply incremental approximation logic to the model. It means that we aim at partitioning the model into a set of fine-grained components to ease the process of incremental approximation. In this regard, we propose two metrics to evaluate which system policy results in the best partitions according to the size and the number of generated components. We categorize the policies into available and unavailable subsets and elaborate a hierarchy of elimination for available policies, indicating how we can achieve the best partitioning policy. In this paper, we investigate the metrics both theoretically and experimentally. The evaluation of the metrics is calculated using an offline approach, and the outcome is deployed at runtime.

\section{System Model and Problem Definition}
The autonomous systems are affected by the changes coming mainly from the environment. As the model is variable, it is updated at runtime, which is composed of two independent parts: environmental and local parts. The environmental part works under uncertainty, and it may face a few changes during its operation. The local part must adjust its behavior to keep the systems' functionalities above an acceptable threshold. Hence, the system needs to follow up the model, be informed about the latest changes, and repeatedly decide about triggering an adaptation action in responding to changes. In a model-driven approach, the model of the system is composed of both local and environmental parts. The local system model needs non-determinism to reflect the choice among actions, and the environment requires probability to describe uncertainty. The best model for describing this type of autonomous system requires deploying a pMDP that provides sequential decision-making among actions and models the changes via a set of parameters.

\begin{definition}\label{def:mdp}
A pMDP is denoted by a tuple $M = (S,T,V,R,P)$ where $S=(s_1,s_2,..., s_n)$ is the finite set of states, $A=(a_1,a_2,..., a_n)$ is the set of actions, $V$ is a finite set of variables,  $P:S \times A \times S \rightarrow \mathcal{F}v$ is the parametric transition probability function that evaluates each parameter $v\in V$, and maps each transition to a real number in [0,1] and $R:S\rightarrow\mathbb{R}$ is the reward function that maps each state to a real number. $\Box$ 
\end{definition}
The autonomous system verifies the model against a few properties at runtime. However, the efficiency of the verification is affected by the large state space of the model. A solution to this issue is to partition the model into a set of components $C=(c_1,c_2,..., c_n)$. Using an incremental approach~\cite{man1}, the components, e.g. Strongly Connected Components (SCCs), can be analyzed and formally verified independently. When a few components are affected by the changes, we need to only re-verify the affected components. Choosing the best partitioning policy leads to efficient verification, and consequently, efficient decision-making mechanism in autonomous systems. We define a policy by Definition~\ref{def:policy}.

\begin{definition}\label{def:policy} A policy $\pi$ for an MDP $M$ is defined as a function $\pi : S \rightarrow {P}(A)$ where {P}(A) is the set of probability distributions on $A$ so that, a policy $\pi$ is a mapping function from any given state s$\in$S to a distribution over actions. $\Box$ 
\end{definition}
In each state $s$ with the available actions $a,b,c\in$A, a policy $\pi$ is formulated as 
$\pi(s): p(a) \times a :+  p(b) \times b :+ p(c) \times c$, where $p(a)+ p(b)+ p(c) = 1$ and $+$ operator denotes a choice. There are two types of policy selection criteria~\cite{policy}: the first approach is based on the probability distribution of actions, and the second is choosing a certain action among available actions. In this research, we mostly follow the second approach because in each situation, the autonomous system must strongly resolve the non-determinism among available actions by applying a policy. 

The research's main problem is finding the best policy for partitioning the model into a set of components. On the one hand, the policy must fulfill the formal requirements of the system. On the other hand, the policy must lead the system toward the best partitioning model in which the outcome includes a set of fine-grained components (see Section~\ref{sec:theory}).

\section{Theoretical Foundations}\label{sec:theory}
An autonomous system may have countless policies to run, and under different situations, it decides which policy should be selected. However, the system cannot choose all possible policies because of environmental changes or internal system constraints limitations. In each situation, a subset of policies cannot be applied. We divide the policies into two categories, including available ($\Pi_{a}$) and unavailable ($\Pi_{u}$) policies. Eliminating unavailable policies makes a subset of states unreachable~\cite{abate3} where some parts of the model are pruned in each situation. This has two effects on the efficiency of the model: the first concerns the state-space reduction of the model, and the second is about eliminating a subset of transitions that leads to forming more independent components~\cite{incremental2}. In this section, we investigate and formulate the theoretical foundations for selecting the best policy among the available policies. The outcome is expected to increase the number of fine-grained components. To assist the autonomous system in deciding the best available policy, we propose two quantitative metrics.

The policies of the system are categorized into available ($\Pi_a$) and unavailable ($\Pi_u$) subsets in which $\Pi = \Pi_a~\cup~\Pi_u$ is hold. In each situation, we aim at pruning the model by eliminating $\Pi_u$ from the choices, and deciding the best policy from $\Pi_a$. The pruning process includes two steps. At first, we eliminate the unavailable policies. As a result of this action, the total state of the system $S$ which consists of reachable $S_a$ and unreachable states $S_{u1}$ ($S = S_a \cup S_{u1}$), changes into a new subset of $S$ represented by $S'= S - S_{u1}$. Similarly, unavailable transitions ($T_u$) are removed from the total transitions $T$ denoted by $T' = T - T_{u1}$. In the second step, a specific policy ($\pi_{i}\in \Pi_{a}$) is selected based on the partitioning results. This time, a subset of $S'$, namely $S_{u2}$, is unreachable, and the set of reachable states is changed into $S" = S' - S_{u2}$. Likewise, the set of transitions $T'$ is changed into  $T'' =T' - T_{u2}$. 

Consider the model $M$ with state-space $S$ is partitioned into a set of components $C$ (e.g. the set of SCCs of $M$). At the first elimination round, $S\rightarrow S'$ is applied, and the model is partitioned into a new set of components $C'$. In the second round of elimination, the model is divided into components $C''$ that result from applying a specific policy as $S'\rightarrow S''$. We expect that $C''$ includes the best available partition of the model in terms of size and number. To define the theoretical criteria for deciding the best policy, we propose two metrics: Balancing and Variation. 

\noindent \textbf{Balancing.} This metric evaluates the effect of policy on the components that considers the ratio of multi-state components' distribution to the number of total components. As denoted in~(\ref{eq:bal}), we show the metric using $Bal(C)$ where it gets the components resulting by applying a policy $\pi_{i}\in \Pi_{a}$, and returns a quantitative evaluation that reveals how much the components are balanced in terms of size. $max$ denotes the maximum number of states in the components, $i$ represents the number of states in a component, and $|C_i|$ indicates how many components with $i$ states exist. As denoted in the denominator, the calculation starts by $i=2$ because the formula does not point to single-state components. As the number of single-state components decreases, the system calculates lower values for $Bal(C)$. The Balancing metric evaluates the effect of policy on the components locally via inter-component comparison.  

\begin{equation}\label{eq:bal}
    Bal(C)=\frac{\sum_{i=1}^{max} |C_i|}{\sum_{i=2}^{max} \frac{1}{|(max - i) + 1|} \times |C_i|}
\end{equation}

\noindent \textbf{Variation} The second metric extends the ability of the autonomous system to evaluate the effect of policy to broader aspects. Balancing only focuses on the size, but Variation globalizes the analysis to the parameters of the components. In our proposed configuration, the autonomous system tends to use incremental verification and reuses the previous results of the components in the next verification steps if those components are not affected by the changes. In other words, as the variety of parameters in a component decreases, the system experiences more efficient verification against the changes. As the value of $Var(C,\theta(V))$ decreases, the components are more resilient against the changes, and the Variation among the components is lower. Formula (\ref{eq:var}) denotes the criteria for calculating the Variation among valuation $\theta$ for a set of parameters $V$ including $(p_1, \cdots,p_n)$. In all parts of the formula, $|C_{val}|$ denotes the number of components that are affected by the changes in the worst-case scenario.

\begin{equation}\label{eq:var}
Var(C, \theta(V))=\frac{\sum_{i=1}^{n} [p_{i}\times |C_{i}|] + \sum_{i=1}^{n-1} [(p_{i}+p_{i+1})\times |C_{(i,i+1)}|] +\cdots + \sum_{i=1}^{n} [p_{i}]\times |C_{(1,\cdots,n)}|} {\sum_{i=1}^{n} p_i\times \sum_{i=1}^{max} |C_i|}
\end{equation}

We argue that the autonomous system finds the best partitioning policy among the set of available policies by calculating Balancing and Variation metrics. The smallest additive value determines the best partitioning policy $\pi_{best}\in\Pi_{a}$. To this end, we propose Lemma~\ref{lemma}, and prove its correctness in Appendix I. 

\begin{lemma}\label{lemma}
The additive value of Balancing and Variation determines the best partitioning policy. $\Box$ 
\end{lemma}

\begin{figure}[h]
\centering
\includegraphics[width=\textwidth]{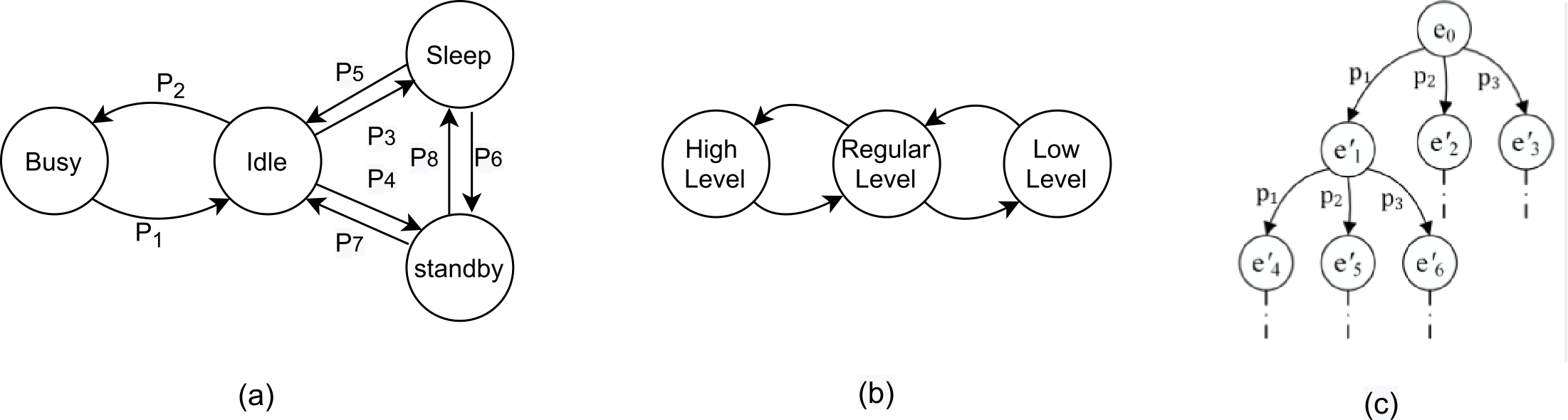}
\caption{The parts of the autonomous system model: (a) sensors, (b) battery model, and (c) the model of the environment. All states of sensors and battery models have self-loop which are skipped in the figure.}\label{fig:model}
\end{figure}

\section{Finding the Best Policy in Practice}\label{sec:res}
To investigate the theoretical findings in practice, we use a case study on an energy harvesting system, propose the steps toward reaching the best partitioning policy and measure the results in practice.

\subsection{Case Study on Self-Adaptive Energy Harvesting Systems}\label{sec:case}
The case study is a self-adaptive solar energy harvesting system which is consists of environmental and local parts. The environment model is responsible for capturing how much energy can be harvested as denoted in Figure~\ref{fig:model}-(c) in which each state determines the expected energy harvesting on an hourly basis~\cite{man3}. The local part is a sensor network with a central battery to save the harvested energy denoted by Figure~\ref{fig:model}-(a) and (b) respectively. Each sensor can work under four operational modes: busy, idle, standby, and sleep, so that the self-adaptive system controls the operating mode to balance the harvested energy and the battery level. We use the PRISM model-checker for modeling and verification~\cite{prism}, and the repository including the codes, models, and sample results is accessible via~\cite{repo}. As mentioned, we use a pMDP to model the autonomous system. If any change occurs, the parameters of the model capture it.

\subsection{The Proposed Solution}\label{sec:sys}

\begin{figure}[h]
\begin{center}
\includegraphics[width=\textwidth]{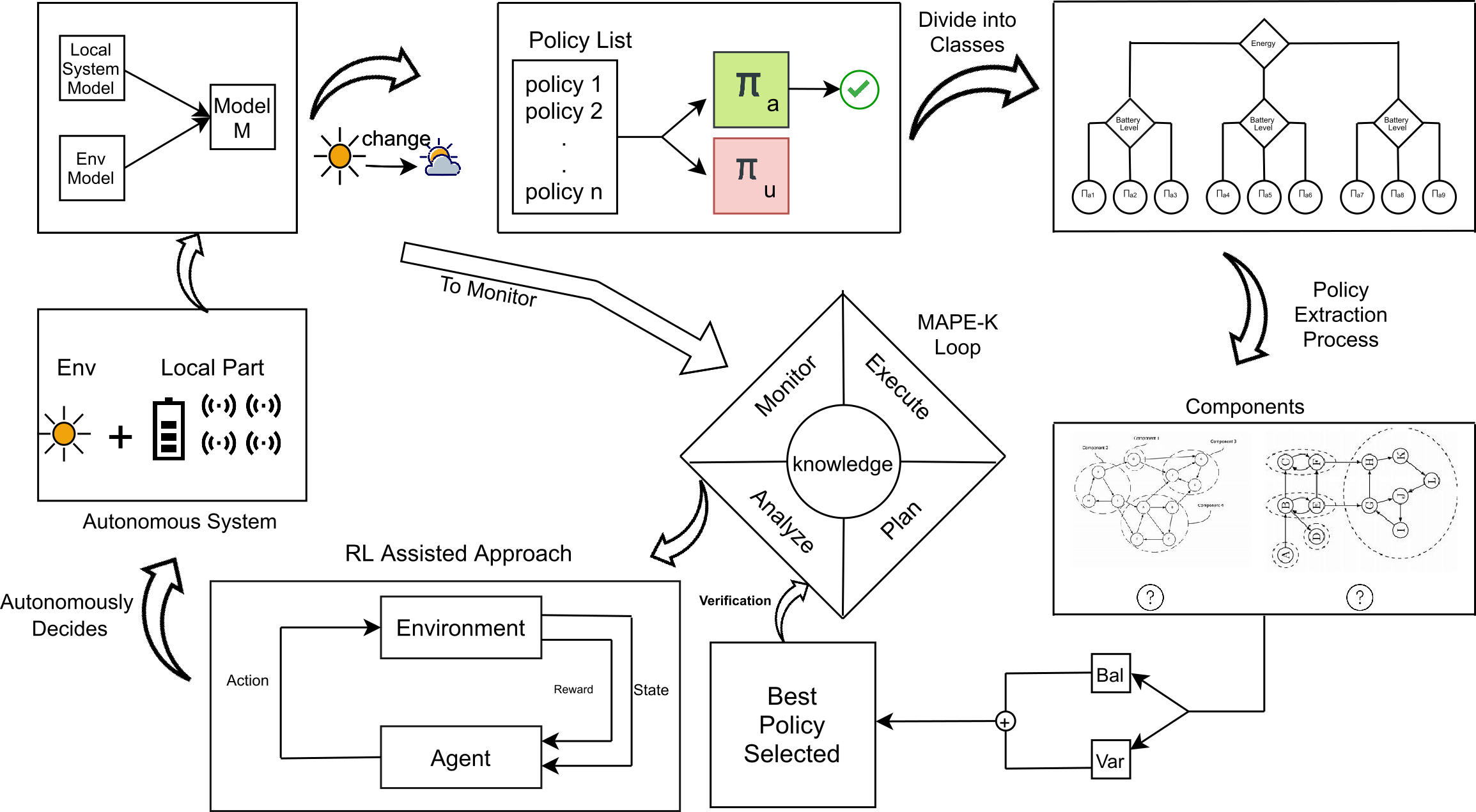}
\caption{The process of calculating the best partitioning policy under each situation of the autonomous system.}\label{fig:system}
\end{center}
\end{figure}

The overview of the proposed system is denoted by Figure~\ref{fig:system}. The system starts by constructing the autonomous system model from the monitoring step and collaborates with MAPE-K to collect information. Then, the system filters available policies $\Pi_{a}$, and constructs a hierarchy of policies regarding the environmental situations and battery levels. The nine categories of policies are investigated to find the best partitioning policy under each situation. Afterward, we analyze the generated components from executing different types of policies. The components are evaluated by Balancing and Variation metrics, and the best (lowest) additive results in each category determine the best policy. As an extension of this process, we plan to use the best policies for training a reinforcement learning-assisted component for decisions beyond these known categories. Note that the entire process of analysis is performed once. However, the autonomous system will deploy the results at runtime. For instance, if the harvested energy is under 200 watt-hours and the battery level is low, the sensors are only allowed to operate through standby and sleep modes, and the best previously investigated policy is selected.

\begin{table}[h]
\scriptsize
\centering
\caption{The results of evaluations for finding the best partitioning policies by the metrics.\label{tab:res}}
\begin{tabular}{!{}l!{}l!{}l!{}l!{}l!{}l!{}!{}l!{}l!{}l!{}l!{}l!{}} 
\hline
    $\pi_i$ &	$p_{2}, p_{3}, p_{4}, p_0$ &	$p_5, p_6, p_0$ &	$p_7, p_8, p_0$ &	$p_2, p_3, p_4, p_0$ &	$p_5, p_6, p_0$ &	$p_7, p_8, p_0$ &	$\#C$ & $\#SS$ & $S:\#C$ & $Bal+Var$
 \\ 
\hline
$b_1$&	0,0.8,0.2,0&	0,0.7,0.3&	0,0.7,0.3&	0,0.5,0.5,0&	0,1,0&	0,0,1&	1584&	1056&	2:528&	4.26\\
$w_1$&	0,0.5,0.5,0&	0,1,0&	0,0,1&	0,0.5,0.5,0&	0,1,0&	0,0,1&	2112&	2112&	1:2112&	infinite \\
\hline

$b_2$&	0,0.8,0.2,0&	0,0.5,0.5&	0,0,1&	0.5,0,0,0.5&	1,0,0&	1,0,0&	528&	0&	2:528&	2.33\\
$w_2$&	0,0.5,0.5,0&	0,0.8,0.2&	0,0.9,0.1&	1,0,0,0&	0.9,0.1,0&	0.9,0,0.1&	396&	0&	2:264&	5.115\\
&	&	&	&	&	&	&	&	&	
4:132&	\\
\hline

$b_3$&	0.9,0,0,0.1&	1,0,0&	1,0,0&	0.2,0,0,0.8&	1,0,0&	1,0,0&	66&	0&	4:66&	4.31\\
$w_3$&	1,0,0,0&	1,0,0&	1,0,0&	1,0,0,0&	1,0,0&	1,0,0&	66&	0&	8:66&	8.33\\
\hline

$b_4$&	0,0,1,0&	0,0,1&	0,1,0&	0,0.7,0.3,0&	0,0.8,0.2&	0,1,0&	6672&	4448&	2:2224&	4.316\\
$w_4$&	0,0.5,0.5,0&	0,0.8,0.2&	0,0.9,0.1&	0,0,0.3,0.7&	0,0.8,0.2&	0,1,0&	5004&	2224&	2:2224&	7.227\\
&	&	&	&	&	&	&	&	&	
4:556&	\\
\hline

$b_5$&	1,0,0,0&	1,0,0&	1,0,0&	0,0.7,0.3,0&	0.1,0.8,0.1&	0.1,0.8,0.1&	1112&	0&	2:556&	1.976\\
&	&	&	&	&	&	&	&	&	
6:556&	\\
$w_5$&	0.8,0,0,0.2&	1,0,0&	1,0,0&	0,0.7,0.3,0&	0,0.8,0.2&	0,1,0&	1668&	0&	2:1112 &	2.117\\
&	&	&	&	&	&	&	&	&	
4:556&	\\
\hline

$b_6$&	0.5,0,0,0.5&	1,0,0&	1,0,0&	0.9,0,0,0.1&	1,0,0&	1,0,0&	556&	0&	4:556&	1.32 \\
$w_6$&	0,0,0,1&	0.9,0,0.1&	1,0,0&	0.4,0.1,0.1,0.4&	1,0,0&	1,0,0&	1112&	0&	4:1112&	1.34\\
\hline

$b_7$&	0.1,0.8,0,0.1&	0,0.5,0.5&	0,0.5,0.5&	0.1,0.7,0.1,0.1&	0,0.8,0.2&	0,0.9,0.1&	2224&	0&	4:1112&	1.983\\
&	&	&	&	&	&	&	&	&	
8:1112&	\\
$w_7$&	0,0.5,0.5,0&	0,0.5,0.5&	0,0.5,0.5&	0,0.5,0.5,0&	0,0.5,0.5&	0,0.5,0.5&	5004&	1112&	2:2224&	6.036\\
&	&	&	&	&	&	&	&	&	
4:1390 &	\\
&	&	&	&	&	&	&	&	&	
8:278&	\\
\hline

$b_8$&	0.9,0,0,0.1&	0.8,0.1,0.1&	1,0,0&	0.1,0.7,0.1,0.1&	0,0.8,0.2&	0,0.9,0.1&	1112&	0&	4:556& 	1.956\\
&	&	&	&	&	&	&	&	&	
8:556&	\\
$w_8$&	0.5,0,0,0.5&	1,0,0&	1,0,0&	0,0,0.5,0.5&	0,0.8,0.2&	0,0.9,0.1&	1668&	0&	2:556&	3.461\\
&	&	&	&	&	&	&	&	&	
4:834&	\\
&	&	&	&	&	&	&	&	&	
8:278&	\\
\hline

$b_9$&	0.5,0,0,0.5&	1,0,0&	1,0,0&	1,0,0,0&	1,0,0&	1,0,0&	278&	0&	8:278&	1.27\\
$w_9$&	0.5,0,0,0.5&	1,0,0&	1,0,0&	0.5,0,0,0.5&	1,0,0&	1,0,0&	556&	0&	4:278&	1.97\\
&	&	&	&	&	&	&	&	&	
8:278&	\\
\hline
\end{tabular}
\end{table}

\subsection{Analytical Results}\label{sec:results}
Table~\ref{tab:res} presents the quantitative results of applying the proposed system to the case study on the energy harvesting system. As denoted in the table, we have reported two evaluations for each subset of available policies, including $b_i$ and $w_i$ that are representative for best and worst analyzed policy in terms of the additive value of metrics that $ Bal+Var$ denotes. The valuation of the parameters for the two sensors is reported. In addition, other effective specifications of the system model are included in the results, such as the number of components $\#C$, number of single states $\#SS$, and number of states in each component $S:\#C$. For parameters' evaluation, we mostly used 0 and 1 policy; however, in some positions, we have used the distribution of the parameters and granted the ability to choose to the system because, in those situations, the valuation of a few parameters do not affect the size of the components. As denoted in the results, Lemma~\ref{lemma} is supported in practice in which the best partitioning policies correspond to the lower additive values of the metrics, and we can intuitively observe that better components correspond to the better additive values of the metrics.

\section{Conclusion and Future Direction}\label{sec:con}
In this research, we tackled the problem of efficient decision-making in autonomous systems. We have used the incremental approximation approach as our base model and proposed a new method for best partitioning the model into a set of independent components. In this regard, we proposed a hierarchy of policies by determining the available policies and then investigating the best partitioning policy among available ones. To evaluate which policy better partitions the model, we have proposed two metrics, including Balancing and Variation, and argue that the additive value of these metrics determines the best partitioning policy. We have investigated the proposed approach both theoretically and experimentally. Future work has already started integrating the proposed system with a reinforcement learning-assisted subsystem to extend it to similar applications where the autonomous systems may face an unprecedented situation.  

\subsection*{Appendix I: Proof of the Lemma}\label{sec:app}
The best partitioning policy depends on two aspects of the components: the first is the size of the components, and the second is the number of generated components. The first is fulfilled if the formula~\ref{eq:bal} guarantees optimization of balancing. So, we discuss and prove how this equation minimizes the quantitative measurement for balancing. Consider that the system includes partitions $C_1,\cdots ,C_{max}$. $max$ denotes the maximum size of the components. We divide the total number of components into a cumulative number of this sequence. By looking at the formula, we can find that the best balancing value is measured when all components have the same size where the minimum value becomes 1. Any other scenario increases the outcome upper than 1. Therefore, it is acceptable that the best value of balancing is 1, and it is the minimum possible calculation. We tend to generate larger components as much as possible. Then, as the size of the components becomes larger, we expect lower values for balancing metrics because the overhead of approximation is decreased. It is reflected in the formula by a coefficient measured in the range of $[0,1]$. The minimum value of coefficients belongs to the single-state components, and one is assigned to $C_{max}$. It can be shown mathematically that the additive numbers of components produce an arithmetic sequence. The validity of Equation~\ref{eq:bal} can be easily proved by induction. The range of this metric is calculated between 1 and infinity, where infinity is the worst case and rarely happens due to the experimental results.

Variation has a different view to the components, which is more globally by investigating the impact of each component regarding the changes in their probabilistic variables. A different number of probabilistic variables in a component makes the system more vulnerable to the changes so that more verification results must be re-evaluated. The minimum number can be achieved when the components do not have any probabilistic variable, and the outcome would be 0. In other words, the maximum value measured by Equation~\ref{eq:var} is 1 when all components contain all parametric variables that are rarely (never) happen. The validity of this equation can also be validated by induction. 

The best policy regarding the metrics is achieved when both generate minimum values, 1 and 0 for Balancing and Variation, respectively. As the metrics work in different ranges of values, we scale Variation by 10. It means the possible range of values for this metric is scaled in the range of $[0,10]$. As mentioned in Lemma~\ref{lemma}, the additive value of these metrics determines the best partitioning policy that is evaluated by minimum value.

\end{document}